\documentstyle[12pt,epsf]{article}
\def\ba{\begin{eqnarray}\samepage}
\def\ea{\end{eqnarray}}
\input amssym.def
\input amssym.tex

\def\double{\Bbb}

\def\rr{{\double R}}

\newcommand{\news}{\setcounter{equation}{0}}
\newcommand{\diff}{\partial}

\newcommand{\be}{\begin{equation}}
\newcommand{\ee}{\end{equation}}
\newcommand{\ben}{\begin{eqnarray}\displaystyle}
\newcommand{\een}{\end{eqnarray}}
\newcommand{\refb}[1]{(\ref{#1})}

\def\Bbb#1{\fam\msbfam#1}

\begin{document}
\title{The Non-Abelian Target Space Duals of Taub-NUT Space}

\author{S.F. Hewson\thanks{email sfh10@damtp.cam.ac.uk}
\\D.A.M.T.P.\\University of
Cambridge\\Silver Street\\Cambridge CB3 9EW\\U.K.}

\maketitle

\begin{abstract}

  We discuss the non-abelian duality procedure for groups
which do not act freely. As an example
we consider Taub-NUT space, which has the local isometry group $SU(2) \otimes
U(1)$. We dualise over the entire symmetry group as well as
the subgroups $SO(3)$ and $U(1)$, presenting unusual new solutions to low
energy string theory. The solutions obtained highlight the relationship between
fixed points of an isometry in one solution and singular points in
another. We also find the interesting
results that, in this case, the $U(1)$ and $SO(3)$ $T$-duality procedures
commute
with each other, and that the extreme points of the $O(1,1)$ duality
group for the time translations have special significance under the $SO(3)$
T-duality.

\end{abstract}

\newpage \eject \renewcommand{\thepage}{ } \pagebreak

\renewcommand{\thepage}{\arabic{page}}

\pagestyle{plain}
\section{Introduction}
\news

The subject of target space duality, or T-duality, in string theory
has generated much interest in recent years. T-duality provides a
method, valid to all orders in perturbation theory, for relating
seemingly inequivalent string theories.

T-duality was originally discovered for the case of a string theory
defined via a $\sigma$-model with an abelian isometry (for example
invariance under time translation). The isometry was then gauged,
leading to another $\sigma$-model describing a different spacetime
geometry. Although the resulting geometry is different from a particle
point of view, the models classically describe the same string theory,
due to the invariance of the generating function under such a
transformation. Quantum equivalence of the two structures is achieved by an
appropriate
shift in the dilaton.

This procedure was then generalised to models with more than an abelian
isometry group of dimension greater than one. This led to the idea of the
vacuum
moduli space; for example, a heterotic string theory with $d$ abelian
isometries and $p$ abelian
gauge fields has an $O(d,d+p)$ invariance. These sort of results and
their implications have been studied in great detail for the
various types of superstring. For a review see, for example, \cite{gpr}.

It was noticed, however, that the basic duality procedure could be generalised
to the case where the original $\sigma$-model had a {\it non-abelian}
group of isometries \cite{qo}. Further work in this direction was
carrried out in \cite{egr} \cite{yl} \cite{grv} \cite{t}. Due to the extra
complexity of non-abelian isometry
groups, the technical problems associated with T-duality in this
framework have only recently been fully solved, in particular relating to the
quantum
equivalence of two string theories, $\cite{steve}$. Non-abelian
duality is thus a valid
symmetry of the string theory, as in the abelian case. It is therefore
of great interest to study new solutions found by application of the
non-abelian
duality formulation. The procedure yields new and unusual solutions, thus
increasing our understanding of low
energy string propagation. One particularly interesting result is that
singular geometries can be dual to regular geometries, from a string
theory point of view. Thus duality
may shed light on the role of spacetime
singularities in string theory.

\medskip

The non-abelian duality shares many
features with the abelian counterpart, but there are also many
differences; for example, in the non-abelian case, the dual model
generally has fewer
isometries than the original model. This means that there is no
immediate way of performing the inverse duality transformation (Hence
`duality' is perhaps a misnomer). The two models
{\it are} equivalent as string theories however, so there should be
some way of defining a duality transformation independent of the
existence of isometries. This implies that for a given $\sigma$-model
there should be a much larger equivalence class of solutions in the moduli
space than the one given by the
abelian duality transformations. A method of performing the reverse
transformation and constucting the full equivalence class of solutions
in the string moduli space
is suggested in \cite{ks} by embedding the string solutions into an
enlarged class of backgrounds, using Drinfeld doubles. For recent
progress on the subject see $\cite{tdrin} \cite{grp}$. Unfortunately \cite{ks}
only consider
{\it{classical}} string equivalence for the case of {\it{freely
acting}} groups. Although interesting, this does not include
cases of groups such as SO(3), which are of fundamental importance in
physics.

\medskip

In this paper we clarify the general method for constructing
the dual models from a $\sigma$-model point of view and present an explicit
calculation of the non-abelian duals
of Taub-NUT space. The resulting metrics are rather unusual and have
entirely non-trivial dilatons and B-fields. These solutions show that the
abelian and non-abelian dualities procedures commute with each other
and single out the extreme $R\rightarrow {1\over R}$ points of
the abelian T-duality as special solutions.

\section{Low Energy String $\sigma$-Models}
\news

$T$-Duality
is an exact symmetry order by order in perturbation theory, so to relate
accurately solutions we need to consider low energy string
propagation. This can be thought of in two ways: either we consider a
string propagating on a massless background or we integrate out the
massive modes of the string in the vacuum.
A bosonic string propagating on a non-trivial background can be
described by a general non-linear $\sigma$-model defined on a
two-dimensional curved surface, because of the correspondence between
the massless modes of a string and a spacetime.

In order that the quantum system is equivalent to the classical system
the model is required to be conformally invariant. This is described by the
non-linear $\sigma$-model in $D$ spacetime dimensions:-

\be \label{nlsm}
S_{nlsm}={1\over{4\pi\alpha'}}\int
d^2\xi\left({\sqrt \gamma}\gamma^{ab}g_{MN}(X)\partial_a
X^M\partial_bX^N +\epsilon^{ab}B_{MN}(X)\partial_a
X^M\partial_bX^N \right) \,
\ee

where $M,N\dots$\ run from $0\dots D-1$ and   $a,b\dots$ run over $  0,1$ \
with $\epsilon^{01}=1$

 $\xi ^a$ are the string worldsheet coordinates and $X^{M}$ are
coordinates in spacetime.
$\gamma$ is the induced metric on the string worldsheet, $\Sigma$. The
background spacetime metric and antisymmetric tensor fields
in which the string propagates are denoted by $g$ and $B$ respectively.

The action has the invariance property

\be \label{ci}\gamma_{ab}\rightarrow e^{\phi(\xi)}\gamma_{ab} \,.
\ee
 This is the conformal transformation of the metric
$\gamma_{ab}$ under which the action remains invariant, since $\gamma$ is two
dimensional.

 Since we are considering bosonic string theory, there is only one more
massless degree of  freedom of the
string, namely the dilaton $\Phi$.
This gives a contribution to the action of the form

\ben \label{dil}
S_{dil} = -{1\over{8\pi}}\int d^2\xi{\sqrt \gamma}R^{(2)}\Phi (X) ,
\een
where $R^{(2)}$ is the scalar curvature of the metric $\gamma$.

This term breaks Weyl invariance on a classical level as do the one
loop corrections to $g$ and $B$.

 Conformal invariance is restored to first order in $\alpha'$ provided that the
string
$\beta$-function equations are satisfied \cite{cfmp}

\ben \label{betafns}
R_{MN}-\nabla_M\nabla_N\Phi -{1\over 4}H_M^{\ LP}H_{NLP}&=&0 \\
\nabla_LH^L_{\ MN}+ (\nabla_L\Phi)H^L_{\ MN}&=&0 \\ \label{beta3}
R-{{D-26+c}\over{3\alpha'}}-(\nabla\Phi)^2 - 2\Box\Phi - {1\over
12}H_{MNP}H^{MNP}&=&0
\, .
\een
where $H=3dB$ and $c$ is a central charge.

It is interesting to note that in the field theoretic limit, the $\beta$
function equations may be considered to be the equations of motion
derived from the low energy effective action
\be\label{leea}
S=\frac{1}{2\kappa^2}\int
d^Dx\sqrt{g}e^{\phi}\left(R-D_{\mu}\phi D^{\mu}\phi+\frac{1}{12}H^2\right)\,.
\ee
In order to agree with the form of the standard general relativity action we
may define the string metric to be related to the Einstein metric by

\be\label{stringmetric}
g_E=e^{\phi}g_S\,.
\ee

 Note that any vacuum solution of the Einstein equations is a solution to
the low energy string theory with zero dilaton and $B$-field, when
tensored with an appropriate conformal field theory to provide a central
charge to satisfy the third $\beta$-function equation
\ref{beta3}. We thus have a good supply of well known solutions to
investigate T-duality.

Finally we remark that all parts of the formalism may be made
supersymmetric. Thus all the results hold in the supersymmetric case.

\section{Gauging of the $\sigma$-Model}
\news

We now consider constucting the dual models from a given string background.
{}From a
spacetime point of view we may perform a duality procedure whenever the string
background
has a group of isometries. We first approach the problem for an action written
in a coordinate
basis, as originally
considered in \cite{qo}, for a background with an isometry group acting without
isotropy, and then look at
the specific case of a symmetry group acting with isotropy. Isotropic
symmetry groups are more difficult to treat due to residual gauge
freedom remaining for any given gauge choice.

\subsection{Isometry Group Acting without Isotropy}

 We consider the low energy bosonic $\sigma$-model in terms of all the
massless modes, and use conformal invariance and diffeomorphism
invariance to put the metric
$\gamma$ into flat form to obtain the action

\ben \label{scoord}
S[X] &=& {1\over4\pi\alpha'}\int {d^2z\Big(
Q_{\mu\nu}{\partial}X^\mu\bar{\partial}X^\nu -
{\frac{\alpha'}{2}}R^{(2)}\Phi\Big)} \nonumber \\
Q &=& g + B \, .
\een
$\mu,\nu$ run from $0 \dots d-1$, where $d$ is the dimension of the
background spacetime we are considering (neglecting compactified
dimensions, for example).
We consider $B$ as a potential of the three form $H$

\be
H=3dB
\ee

\medskip

 Suppose that $Q$ has a freely acting group of isometries ${G}$. This means
that there exists
Killing vector fields ${\cal T}^i$ (the generators of $G$) such that

\ben
{\cal L}_{\cal T}g&=&0\nonumber\\
{\cal L}_{\cal T}\Phi&=&0\nonumber\\
 {\cal L}_{\cal T}H&=&0 \nonumber \\
\Rightarrow {\cal L}_{\cal T}B&=&d\omega \mbox{ \ locally, for some
one-form }\omega\,.
\een

Note that since the commutator of two Killing vectors is also a
Killing vector, the Killing vectors can be thought of as defining a
Lie algebra of some group $G$.

We may introduce a gauge field into
the action to describe this symmetry (or a subgroup thereof) by the
minimal coupling prescription:

$$\partial X^m \rightarrow DX^m = \partial X^m + A^a
{(T_a)^m}_nX^n$$

 The $T_a $ are the generators of the Lie algebra in the adjoint
representation of the subgroup $H \subset G$ and give
rise to the structure constants ${f^c}_{ab}$

\ben [T_a,T_b] &=& {f^c}_{ab} T_c \nonumber \\
Tr(T_a T_b) &=& 2{\eta}_{ab} \, ,
\een
$\eta$ is the Cartan matrix of the group and $a,b$ are internal group indices
running over 1..dim($H$).
\medskip
 In order to reproduce the original action, the gauge fields must be
constrained to be flat by the addition of the Lagrange multiplier term

\be
S_{constraint} =  {1\over4\pi\alpha'}\int {d^2z\ Tr\left(\Lambda F\right)} \, .
\ee

 The equations of motion for $\Lambda$ give a vanishing, non-dynamical, $F$.

\ben
F &=& \partial {\bar A} - {\bar \partial}A + [A,{\bar A}]=0 \nonumber \\
\smallskip
&\iff& A = h^{-1}\partial h \nonumber \\
&\ & {\bar A} = h^{-1}{\bar \partial}h \nonumber \\
&\ & h \in H \, ,
\een
that is, $A$ is pure gauge.
\smallskip

 As described in the model above the X are the dynamical fields and the
$\Lambda$ are Lagrange multipliers. However, we may also perform an integration
by
parts on the constraint term, which will introduce derivative terms
of $\Lambda$ into the action. Thus the Lagrange multipliers will be promoted to
dynamical
variables and the coordinates describing the isometry become
constant.
\medskip

 The form of the dual is given by

\be
S_{dual} = {1\over4\pi\alpha'}\int {d^2z \left({\bar {h_a}}
\left(f^{ab}\right)^{-1} h_b +Q_{ij}\partial x^i {\bar\partial}x^j \right)} \,
\ee

where
\ben
h_a &=& -\partial \Lambda_a +  Q_{\mu n}\partial X^\mu{(T_a)^n}_m X^m \nonumber
\\
{\bar h}_a &=& \ {\bar\partial} \Lambda_a +  Q_{n\mu}{\bar\partial}
X^\mu{(T_a)^n}_m X^m \nonumber \\
f_{ab} &=& - {f_{ab}}^c \Lambda_c +
X^p{(T_b)^q}_p Q_{qn}{(T_a)^n}_mX^m \, .
\een
where {$m,n,p,q  $ :1,2,... dim$H$}
and {i,j:dim($H$)+1...dim$Q$}.

\medskip
 We now have the dual model from which we can read off the new Q field
in terms of the coordinates on the dual manifold
$(x^i,\Lambda^m)$. Note that if the non-abelian gauged subgroup is
semisimple then the gauging procedure is well defined $\cite{steve}$.

 Classically the two solutions generated are equivalent, and in the
special case where the gauged symmetry group is semisimple then in the
low energy limit the new fields satisfy the beta function equations if we
make the dilaton shift $\cite{buscher}$

\be \Phi_{new} =  \Phi_{old}-\log(\det f) \,.
\ee

 These equations reduce to the abelian $T$-duality equations$\cite{gpr}$
for a group of abelian isometries, in which case for the two models to be truly
dual as string theories the new
coordinates $\Lambda$ must have an appropriate periodicity. The
periodicity is chosen to remove any potential conical singularities in
the spacetime, where the dual would not satisfy the $\beta$-function
equations.

Notice that the new fields will be singular at the points where
det$(f)=0$. This certainly occurs at the fixed points of the
action of the gauged symmetry group $\cite{quev}$. If the symmetry
group gauges is isotropic then there are by definition fixed points of
the action. Hence the duals to such spaces will always have curvature
singularities. This is due to a bad gauge choice where the singularity occurs.

\subsection{Case with Isotropy}

 If the gauged subgroup has an isotropy group, as is often the case for
groups in physics, then there is an additional freedom
in the gauge fixing. If we choose the gauge fixing $X={\hat X}$, then the
theory is still
invariant under
\ben \label{gaugefix} {\hat X}&\rightarrow&L{\hat X} \nonumber \\
\Lambda&\rightarrow&L^{-1}\Lambda L
\een
where $L$ is in the little group of ${\hat X}$ under the action of $G$,
$lg(\hat X)$.
Thus there is a redundancy in the new coordinates $\Lambda$, which
must be fixed, by placing $\dim \left(lg({\hat X})\right)$
constraints on the $\Lambda$.

Note that this extra gauge freedom implies a fixed point in the action
of the group. This implies that the dual models will have curvature
singularities.

{\section{Duality Transformation via Forms}\label{section}}
\news

 The duality procedure becomes clearer if we choose a
non-coordinate basis such that the background metric and antisymmetric
tensor field are constant under the action of the symmetry group. In
the case of a non-isotropic symmetry group the isometry dependent variables can
be factored from
$Q$ in \ref{scoord} a unique, up to constant factors, way and the
action can be rewritten in the form \cite{t}

\ben
S &=& {1\over{4\pi\alpha'}}\int d^2z \Big(e^m{{\cal{E}}}_{mn}(x^i){\bar
e}^n+e^m{{\cal{E}}}_{mj}(x^i){\bar \partial}x^j+\partial
x^i{{\cal{E}}}_{in}(x^i){\bar e}^n\nonumber \\
&\ & \ \ \ \ \ \ \ \ \ \ \ \ \ \ \ \ \ \ \ +\partial x^i
{{\cal{E}}}_{ij}(x^i){\bar\partial}x^j - {\alpha'\over {2}}{{\sqrt
\gamma}}R^{(2)}\Phi \Big) \  \,
\een
\ben\label{mc}
e^m \equiv {e^m}_\mu d x^\mu  \nonumber \\
d{e}^m + {1\over2}{f^m}_{pq} e^p \wedge e^q = 0 \,
\een
where $m$,$n$ are isometry indices and $x^i$ are the other coordinates.

 $e^m$ satisfy the Maurer-Cartan equation \ref{mc}, implying that the
connection $e^m$ is pure
gauge
$$ e^m_\mu(x^\nu)\partial x^\mu = \mbox{Tr}\left({\cal T}^mg^{-1}\partial g
\right)$$
so we may choose $g=1\iff x=0$.

 Equivalently the ${\cal T}_m$ act as Killing vectors on ${\cal{E}}_{\mu\nu}$.

 If we gauge the $\sigma$-model with $G$ valued gauge fields
then in the low energy limit the duality procedure yields
\ben \label{formsmethod}
{\hat S}&=& {1\over{4\pi\alpha'}}\int d^2 z \Big( (\partial
\Lambda_m-\partial
x^i{\cal{E}}_{im})\left[(f_{qq'})^{-1}\right]^{mn}({\bar\partial}
\Lambda_n+{{\cal{E}}}_{nj}{\bar \partial}x^j)\nonumber \\
&\ & \ \ \ \ \ \ \ \ \ \ \ \ \ \ \ \ \ \ \  + {{\cal{E}}}_{ij} \partial x^i
{\bar \partial}x^j- { \alpha'\over
2 }{\sqrt \gamma}R^{(2)}\Phi_{new} \Big)  \nonumber \\
\Phi_{new}&=&\Phi-\log(\det(f))\nonumber \\
f_{qq'}&=&{{\cal{E}}}_{qq'}+\Lambda_p{f^p}_{qq'}\, \
\een

 The original Lagrange multipliers $\Lambda$ have become new
coordinates in the theory, the old isometry coordinates are now
fixed. We can now read off the new $Q$ field in terms of the
coordinates on the dual manifold $(x^i,\Lambda^m)$.

\medskip
If the original solution was an exact conformal field theory
then we may obtain an exact dual conformal field theory by an
adjustment of $\ref{formsmethod}$ to include higher order $\alpha'$
corrections $\cite{steve}$.

\subsection{Isotropic Case}

 If the symmetry group is isotropic then there are extra degrees of
freedom in choosing the decomposition of $Q$ into
a form which has constant backgrounds with respect to the action of
the isometry. For an isotropic symmetry, the dimension of the group $\dim(G)$
is larger then the dimension of the surface of transitivity $\dim(S)$.
$$\dim(G)=\dim(S)+\dim(\mbox{lg}(G))$$
 In order
to perform the duality transformation it is therefore
neccessary to consider the extension of the surface over
which the isometry acts to a
higher dimensional surface of dimensionality $\dim(G)$. We then
choose a particular decomposition of $Q$ and
perform the non-abelian duality transformation using $\ref{formsmethod}$. In
the resulting solution the new coordinates are constrained
to lie on some surface corresponding to the action of the isotropy
group of the original gauge choice. Choosing different points on this
surface merely corresponds to a change of variables in the new
solution.

The action of the isotropy group, $R\subset H$, corresponds to the action

\ben\label{constraint}
f_{qq'}&\rightarrow& R^{-1}f_{qq'}R\nonumber \\
&=&R^{-1}\left({\cal{E}}_{qq'}+\Lambda_p{f^p}_{qq'}\right)R\nonumber\\
&=&{\cal{E}}_{qq'}+{\hat \Lambda}_p{f^p}_{qq'}\,.
\een
Thus we may choose an element $R$ to constrain $\Lambda$ in the above fashion.

\section{Taub-NUT space}
\news

 Taub-NUT is a vacuum solution to the Einstein equations in four dimensions.
          $$ R_{\mu\nu}=0$$
The line element can be written as \cite{he}
\be \label{TN}
ds^2  =  -f_{1}(dt + 2l\cos\theta d\phi)^2 +{1\over f_{1}} dr^2 +(r^2 +l^2)
(d\theta^2 + {\sin\theta}^2 d\phi^2) \, ,
\ee
$$f_1=1-2\left({{mr+l^2}\over{r^2+l^2}}\right)={(r-r_+)(r-r_-)\over(r^2+l^2)}$$
$$r_{\pm}=m\pm\sqrt{(m^2+l^2)}$$
where $l$ is the NUT charge and $m$ is the mass.

 For constant $r$, this is the metric on a squashed three sphere,
and for non-zero $l$ is written as

\ben \label{3sr}
ds^2  &=&  -(2l)^2 f_{1}(d\psi + \cos\theta d\phi)^2 +{1\over f_{1}}
dr^2 +\Omega^2 (d\theta^2 + {\sin\theta}^2 d\phi^2) \nonumber \\
\psi &=& t \over 2l \nonumber \\
\Omega^2&=&r^2+l^2\, .
\een
In order to avoid conical singularities $\psi$ must be identified with
period $4\pi$.

\bigskip
 For constant $r$, each point on the manifold $M$ defined by the line element
can be thought of as representing a
rotation on 3-$d$ Cartesian vector ${\bf r}$, so Taub-NUT represents a
group manifold. The range of $\psi$ determines
the group as $SU(2)$.

 The group elements are

\be
 G(\theta,\psi,\phi)  =  \exp \left({ \psi \sigma_z \over 2i}\right) \exp
\left({ \theta \sigma_x \over 2i}\right) \exp \left({ \phi \sigma_z
\over 2i}\right)\,
\ee
which act on $r$.$\sigma$, where the $\sigma$ are the usual
Pauli spin matrices. In this context $\psi$, $\phi$ and $\theta$ are the Euler
angles of a general
3-$d$ rotation.
 Topologically, $SU(2)$ is the three sphere $S^3$ which admits a Hopf
fibration

\be
S^3\rightarrow S^2:(\psi,\theta,\phi)\rightarrow (\theta,\phi)
\mbox{ with fiber } S^1\,.
\ee

 For Taub-NUT space we have the fibration
\be
\pi : M \rightarrow S^2 \  (r,\psi,\theta,\phi)\rightarrow
(\theta,\phi)\,
\ee
with the metric on the fiber given by \refb{3sr} with
constant $\theta,\phi$.
\be
ds^2 =  -(2l)^2 f_{1}d\psi^2 +{1\over f_{1}}dr^2\,
\ee
this fiber is topologically $R\times S^1$.

 However, Taub-NUT space has the topology $R^1\times S^3$ which is
only locally $(R^1\times S^1)\times S^2$. Thus Taub-NUT has a non-trivial
fibration which cannot be
globally written as a direct product.

\subsection{Geometry of Taub-NUT space}

\begin{figure}
\epsffile{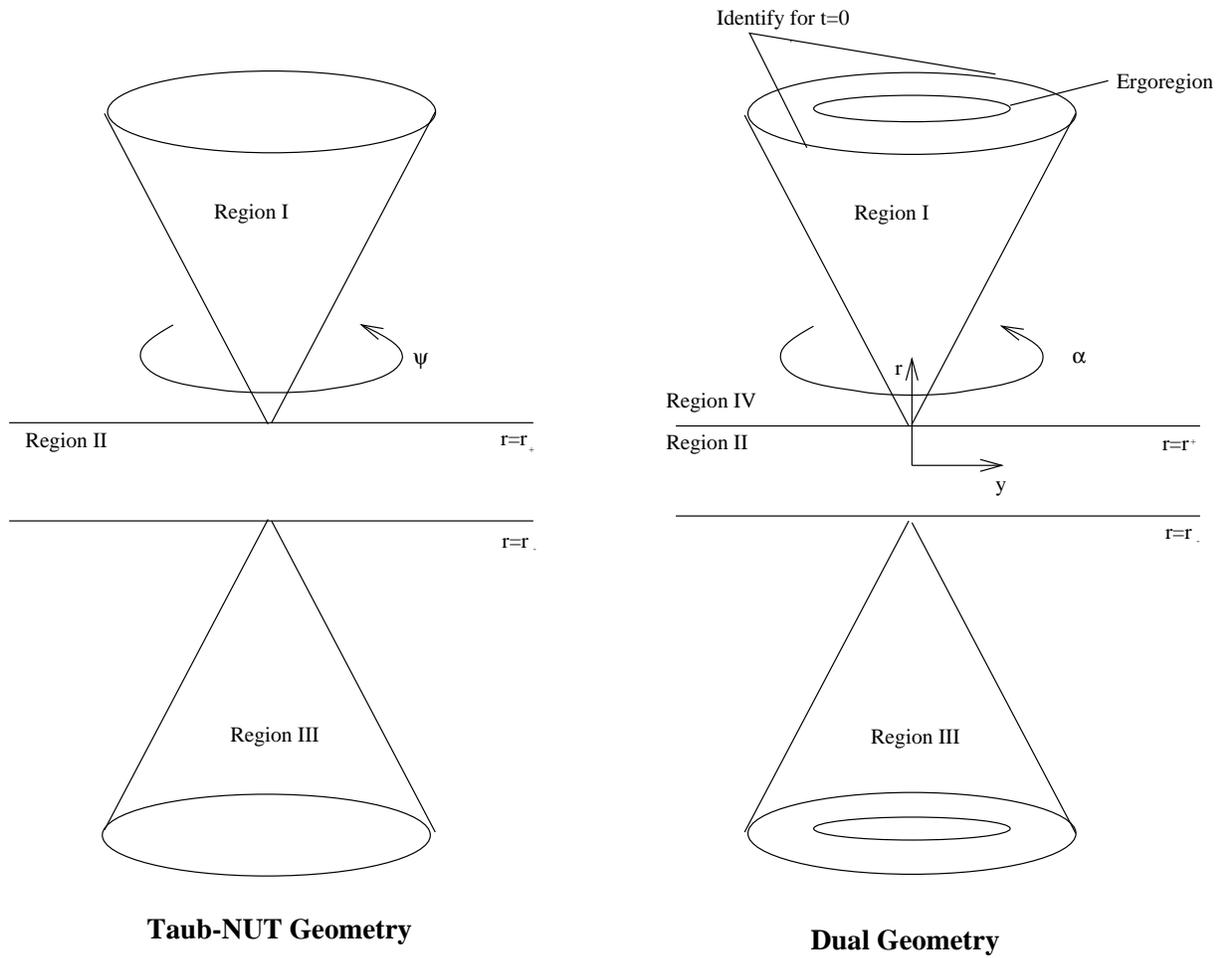}
\caption{Taub-NUT and its dual}
\label{fig1}
\end{figure}

Taub-NUT space is regular with no horizons, but there are still three
separate regions of the spacetime which need to be
considered (See figure \ref{fig1}).

\medskip

{\bf Region I:  $r>r_+$}

 In this region the geometry of Taub-NUT space can be thought of
as that of rotations of a cone in
$\rr^3$ . The cone is fixed at the vertex, corresponding to $r=r_+$, and has an
internal symmetry axis corresponding to
the right action of $\psi$.

 There are closed timelike curves in this region due to the
periodicity of the $\psi$ variable.

\bigskip

{\bf Region II:  $r_-\leq r\leq r_+$}

 This region represents a homogeneous cosmological
model where $r$ acts as the timelike variable. Although this region is
compact and regular there are families of incomplete timelike and null
geodesics
which spiral in towards $r=r_{\pm}$. Thus the region is {\it non-Hausdorff}.

\bigskip
{\bf Region III}

 This has a similar geometry to region I. In the massless case, $m=0$, the two
regions are isometric and may be identified with each
other.

\subsection{Monopole Interpretation of Taub-NUT}

 The non-Hausdorffness of Taub-NUT not too difficult to deal with since we do
have
bifurcation of geodesics. If we consider the euclidean continuation of
Taub-NUT by taking $t\rightarrow it, l\rightarrow ir$, then the resulting
instanton is Hausdorff. Geodesics in this metric
represent asymptotic motion of monopoles \cite{gm}, thus Taub-NUT space has
a clearly defined physical interpretation.

\subsection{Isometries of the Group Manifold}

The Killing vectors for the line element \refb{3sr} act on surfaces
of constant $r$, so we may neglect the $dr$ terms throughout the
discussion of the symmetries.

 For constant r, the line element becomes
\be \label{metric}
ds^2  =  -(2l)^2 f_{1}(d\psi + \cos\theta d\phi)^2 +(r^2 +l^2)(d\theta^2 +
{\sin\theta}^2 d\phi^2) \, .
\ee

 The left invariant one forms for this line element are
\ben \label{TNforms}
e^x&=& \sin\theta \sin\psi d\phi + \cos\psi d\theta \nonumber \\
e^y&=& \sin\theta \cos\psi d\phi - \sin\psi d\theta \nonumber \\
e^z&=& d\psi + \cos\theta d\phi \, .
\een

 This enables us to write the metric in terms of the flat connection
\be
ds^2 = -(2l)^2f_1{e^z}^2 +
(r^2+l^2)({e^x}^2+{e^y}^2) \, ,
\ee
since these are Maurer-Cartan forms obeying the relation
\be
d{e}^m + {1\over2}{f^m}_{pq} e^p \wedge e^q = 0 \,,
\ee
where ${f^m}_{pq}$ are the structure constants of the rotation group:
\be {f^m}_{pq}=-\epsilon^m_{\ pq}\,.
\ee
Equivalently, we may consider the  left invariant vector fields (Killing
vectors)
\ben
L_x &=& \cos\psi {\partial\over{\partial\theta}} - \sin\psi \Big(\cot\theta
{\partial\over{\partial\psi}} -
{1\over \sin\theta}{\partial\over{\partial\phi}} \Big) \nonumber \\
L_y &=& -\sin\psi {\partial\over{\partial\theta}} - \cos\psi \Big(\cot\theta
{\partial\over{\partial\psi}} -
{1\over \sin\theta}{\partial\over{\partial\phi}} \Big) \nonumber \\
L_z &=& {\partial\over{\partial\psi}} \, .
\een
The line element is invariant to first order under the action
$$ x \rightarrow x + \epsilon L $$
and the algebra of these vectors is given by
\ben
\Big[L_i,L_j\Big] &=& -\epsilon_{ijk} L_k \nonumber \\
\epsilon_{123}&=&+1 \, .
\een
This is the same as the Lie Algebra for ordinary rotations on the two
sphere, because the extra $\psi$ coordinate is abelian.

\section {Duality of Taub-NUT}
\news

Although a simple solution of the Einstein equations, Taub-NUT has a
rich duality structure.  There are different ways in which the
solution can be gauged, leading to interesting solutions with
non-trivial $B$-fields and dilatons.
The first, well known as the H-monopole (if one considers the
euclidean continuation of the solution), is the abelian dual of
the solution with respect to $\frac{\diff}{\diff t}$. The H-monopole has an
SO(3) symmetry over which we can
take the T-dual. The second is the result of gauging out the entire
symmetry group of Taub-NUT. The resulting solution has one abelian symmetry.
 We show that the abelian duality procedure commutes with the one for
non-abelian SO(3) duality.

Note that it is also possible to consider the dual for the Killing
vector  $\frac{\diff}{\diff \phi}$. The resulting solution has no
spherical symmetry, and we do not consider the results obtained in
this case.

\subsection{SO(3) Dual of Schwarzschild Solution ($l=0$ case)}

 The Taub-NUT solution reduces to the Schwarzschild solution in the
zero NUT charge limit. The group structure is discontinuous in the
limit, however, since these two solutions have a different
symmetry structure, in that the Schwarzchild solution has a trivial
fibration which can be written globally as a direct product. The
Schwarzschild solution has the isometry group $SO(3)$, the action of
which which has an
${\it isotropy}$ subgroup.

Due to this isotropy of the $SO(3)$ gauge group there are three Killing
vectors which act on a two-dimensional surface of $(\theta,\phi)$,
hence there will be an SO(2) freedom in the new coordinates if we take
the dual of the Schwarzschild solution. We thus consider the solution
as a surface embedded in a larger space, and
perform the duality procedure over {\it three} forms $e^1,e^2$ and
$e^3$, by choosing forms satisfying the Maurer Cartan equation \ref{mc} such
that

\be{\label{decomp}}
d\theta^2+\sin^2\theta d\phi^2=(e^1)^2+(e^2)^2+(e^3)^2\,.
\ee

This choice has an ambiguity defined up to an $SO(2)$ rotation. After
taking the
dual we must therefore fix the extra $SO(2)$ gauge freedom in the new
coordinates $\Lambda$.

 We write the line element in terms of the forms
\ben \label{newforms}
e^1 &=& d\theta \nonumber \\
e^2 &=& \sin\theta d\phi \nonumber \\
e^3 &=& \cos\theta d\phi \, .
\een

 Note that we have made an implicit choice of gauge. We generally may choose
\ben \label{gaugechoice}
e^2 &=& A\sin\theta + B\cos\theta \nonumber \\
e^3 &=& A\cos\theta - B\sin\theta \nonumber \\
& & A^2+B^2 = 1 \, .
\een
In terms of the forms \ref{newforms}, the line element reads:-
\be
ds^2=-\left(1-\frac{2m}{r}\right)dt^2+\left(1-\frac{2m}{r}\right)^{-1}dr^2+
r^2\left( (e^1)^2 + (e^2)^2 \right) \,
\ee

By extending the surface of transitivity of the Killing vectors to a
three-dimensional surface we can now dualise over the $e^1,e^2$ and
$e^3$, treating the problem as for
non-isotropic symmetry groups.
We obtain:-
\be
f_{qq'} = {{\cal{E}}}_{qq'} + \Lambda_p{f^p}_{qq'} =
\pmatrix{r^2&\Lambda_3&-\Lambda_2\cr-\Lambda_3&r^2&\Lambda_1\cr\Lambda_2&-\Lambda_1&0}\,
\ee

 This matrix can be inverted to give the new fields. We must
then
constrain the new coordinates, $\Lambda$, by making a gauge choice which leaves
the original ${\cal{E}}$ field unchanged.
In this case the freedom in the dreibein is given by rotations about $e^3$,

\ben
\pmatrix{{\hat e^1}\cr{\hat
e^2}}&=&\pmatrix{&\cos\Theta&\sin\Theta\cr&-\sin\Theta&\cos\Theta}\pmatrix{{e^1}\cr{e^2}}\nonumber
\\
{\hat e}^3&=&e^3\nonumber \\
\Rightarrow ({\hat e}^1)^2 + ({\hat e}^2)^2 &=&(e^1)^2 + (e^2)^2 \,.
\een

Imposing this constaint as in $\ref{constraint}$ by choosing, for example,
$\Lambda_2=0$, we
obtain the extra contribution to the metric (after relabelling all the
$\Lambda_i$)

\ben
{\hat{g}}&=&\bordermatrix{&\partial\Lambda_1&\partial\Lambda_2\cr&\Lambda_2^2&\Lambda_1\Lambda_2
\cr&\Lambda_1\Lambda_2&(r^4+\Lambda_2^2)}{1\over
\Delta}\nonumber \\
\Delta&=&r^2\Lambda_1^2\,.
\een
Of course, a different constraint choice merely corresponds to a
change of variables in the new solution.

We  now make the change of variables $y=\Lambda_2$ and $x^2+y^2=\Lambda_1^2$,
the solution reduces to
that found in \cite{qo} via coordinate methods
\ben\label{xxxx}
ds^2&=&-\left(1-{2m\over r}\right)dt^2 +\left(1-{2m\over r}\right)^{-1}dr^2
+{1\over
{r^2(x^2-y^2)}}[r^4dy^2+x^2dx^2] \nonumber \\
B&=&0 \nonumber \\
\Phi&=&-\log[r^2(x^2-y^2)]\,
\een

\subsubsection{Analysis of the dual space}

The line element obtained by taking the $SO(3)$ dual of
Schwarzschild space is important to understand, because any space with
spherical symmetry will give rise to a dual space of a similar form.

There are several points of note.

Firstly we note that the ${\it five}$ dimensional space
obtained after performing the duality procedure, but without placing a
constraint on the
$\Lambda_i$, has a singular metric with zero determinant. Thus the dual
solution obtained is truly four dimensional, and
may not be thought of as  just a slice through a larger
five-space string theory solution.

Also, since the duality is performed over
the spherically symmetric part of the metric, the  dual solution
retains the original horizon at $r=2m$ and a curvature
singularity at $r=0$.  New curvature
singularities are introduced along the lines $x=\pm y$, corresponding to the
axes of fixed points of the original symmetry, $\sin\theta=0$. The $SO(3)$
isometry is lost
after the duality transformation, and it has been shown that the
metric only has one continuous isometry corresponding to the abelian time
shift \cite{qo}. The only new symmetries introduced under the duality
procedure are the discrete symmetries $x\rightarrow-x\mbox{ and
}y\rightarrow-y$, and we may choose to identifty these points. To perform the
inverse transformation we need to
consider an enlarged background containing both the original and
dual spacetimes \cite{ks}.

More properties of the original spacetime, other than the isometries,
are also lost after dualising. The resulting spacetime is not algebraically
special (Petrov type I) and, in the massless case, does not admit any
Killing spinors.

 Despite such problems, however, the resulting spaces may be
given a qualitative analysis. Such a discussion will be given in
section \ref{proptn}.

\subsection{H-Monopole}

 Standard application of the abelian duality transformation
for the Killing vector ${\diff \over {\diff t}}$ on Taub-NUT yields the
{\it{H-Monopole}}  given by :-

\ben
ds^2 &=& -{1\over {(2l)^2f_1}} d{\Lambda_0}^2 + \Omega^2(
d\theta^2+\sin\theta^2d\phi^2 )+ {1\over f_1} dr^2
\nonumber \\
B_{0 \phi}&=& -\cos\theta \nonumber \\
\Phi &=& - \log{\Delta_0} \nonumber \\
\Delta_0 &=& \det(g_{00}) = -(2l)^2f_1  \, .
\een
\medskip

 This solution now has non-trivial $H=3dB$-field and the line element has an
$SO(3)$ symmetry. Thus we can either perform duality over $\Lambda_0$
which will return us to Taub-NUT space, or we can gauge the $SO(3)$ spherical
symmetry, which will elucidate the method of calculating duals given
in section \ref{section}.

\subsection{ $SO(3)$ dual of H-Monopole}

 We rewrite the H-monopole line element in terms of the
particular choice of
Maurer-Cartan forms :-
\ben
e^0 &=& d\Lambda_0 \nonumber \\
e^1 &=& d\theta \nonumber \\
e^2 &=& \sin\theta d\phi \nonumber \\
e^3 &=& \cos\theta d\phi \, .
\een

 yielding:-
\ben
ds^2 &=& -{1\over {(2l)^2f_1}}(e^0)^2 + {\Omega}^2 \big( (e^1)^2 + (e^2)^2
\big) +
{1\over{f_1}}dr^2 \nonumber \\
B_{0 \phi}&=& -\cos\theta \Rightarrow B = \cos\theta d\Lambda_0 \wedge
d\phi = e^0 \wedge e^3 \nonumber \\
&&\Rightarrow B_{03}=1 \, .
\een

 We now dualise over $e^1,e^2,e^3$ treating the choice as fixed.
We obtain:-
\be
f_{qq'} = {{\cal{E}}}_{qq'} + \Lambda_p{f^p}_{qq'} =
\pmatrix{\Omega^2&\Lambda_3&-\Lambda_2\cr-\Lambda_3&\Omega^2&\Lambda_1\cr\Lambda_2&-\Lambda_1&0}\,
\ee

\ben
\partial\Lambda_m&-&\partial x^i{{\cal{E}}}_{im}:\
\big(\partial\Lambda_1,\partial\Lambda_2,\partial\Lambda_3-\partial\Lambda_0\big)\nonumber
\\
{\bar\partial}\Lambda_m&+&{{\cal{E}}}_{nj}{\bar\partial}x^j:\
\big({\bar\partial}\Lambda_1,{\bar\partial}\Lambda_2,{\bar\partial}\Lambda_3-{\bar\partial}\Lambda_0\big)\nonumber
\een
 Inverting $f$ we read off the new contributions to the
metric and anti-symmetric tensor field. Since the new line element is
in the five dimensional space
$(\Lambda_0,r,\Lambda_1,\Lambda_2,\Lambda_3)$ we must
make a constraint on the new variables
$(\Lambda_1,\Lambda_2,\Lambda_3)$ by considering elements which leave
the gauge choice fixed, as in \ref{gaugefix}. In this case we may
make the particular choice $\Lambda_1=0$. Note that the five space
metric has zero determinant, and actually describes a four
dimensional surface. The extra gauge fixing merely removes the
redundancy in the new coordinates $\Lambda_i$ and all such choices reduce
to the same four-space.

 After imposing
the constraints we find that:-
\ben
\hat{g}&=&\bordermatrix{&\partial\Lambda_0&\partial\Lambda_1&\partial\Lambda_2\cr&\Omega^4+\Lambda_2^2&-\Lambda_1\Lambda_2&-(\Omega^4+\Lambda_2^2)
\cr&-\Lambda_1\Lambda_2&\Lambda_1^2&\Lambda_1\Lambda_2\cr&-(\Omega^4+\Lambda_2^2)&\Lambda_1\Lambda_2&(\Omega^4+\Lambda_2^2)}{1\over
\Delta}\nonumber \\
\mbox{{\bf B}}&=&0 \nonumber \\
\Delta&=&\Omega^2\Lambda_1^2\nonumber \\
\Phi&=&\Phi_{H-monopole}-\log\Delta\nonumber \\
&=&\log f_1-\log(\Omega^2\Lambda_1)\,
\een
 To this metric we need to add the $(\Lambda_0,r)$ sector of
the H-monopole, resulting in the solution we shall call $H_{SO(3)}$:-
\be \label{HSO3}
ds^2=-\frac{d\Lambda_0^2}{(2l)^2f_1}+\frac{dr^2}{f_1}+{1\over
\Delta}(-\Lambda_2d\Lambda_0+\Lambda_1d\Lambda_1+\Lambda_2d\Lambda_2
)^2+{1\over\Delta}\Omega^4(d\Lambda_0-d\Lambda_2)^2\,
\ee
\medskip
Note we have explicitly checked that this background is a solution to
\ref{betafns}
\bigskip

 Clearly this solution has the abelian isometry
$\Lambda_0\rightarrow\Lambda_0+$ constant. We can therefore perform
the $T$-dual for this variable. We consider specifically the extreme
case of $\Lambda_0\rightarrow{1\over\Lambda_0}$ duality, although this
could be extended to an $O(1,1)$ symmetry to generate an entire spectrum of
solutions.

\subsection{Abelian dual of $H_{SO(3)}$}

 Performing the standard $T$-duality and making the variable
change
$${\hat \Lambda_0}=\alpha\, \Lambda_1=(2l)^2y\, \Lambda_2=(2l)^2t$$
we obtain the solution $H_{SO(3),\Lambda_0}$:-

 \ben \label{HSO3T}ds^2&=&-{{\Omega^4}\over\Delta}dt^2 -
{(2l)^2\over{\Delta}}\big(tdt+ydy)^2+{{f_1\Omega^2(2l)^2}\over\Delta}\big(dy^2+y^2d\alpha^2\big)+{dr^2\over
f_1} \nonumber \\
 B&=&\left({{{\Omega^2y^2(2l)}\over\Delta}-(2l)}\right)d\alpha\wedge
dt+{{(2l)^3f_1yt}\over\Delta}d\alpha\wedge dy \nonumber \\
\Phi&=&\log(-(2l)^2\Delta)\nonumber \\
\Delta&=&f_1\big(\Omega^4+(2l)^2t^2\big)-\Omega^2y^2\,.
\een

\subsection {SU(2) Dual of Taub-NUT}

We note that the symmetry structure of Taub-NUT has a non-trivial
decomposition and we rewrite the line element $\ref{TN}$ using the forms
\ref{TNforms}. We are
now able to dualise over the entire SU(2), instead of just an abelian
subgroup. The symmetry acts on surfaces of constant $r$ so we may neglect the
$dr^2$
term throughout the duality procedure. As we cannot decouple the
$\psi$ variable from the symmetry, there is only one abelian degree of
freedom in our choice of tetrad.

 Equations \ref{formsmethod} give:-

$$f_{qq'} = {{\cal{E}}}_{qq'} + \Lambda_p{f^p}_{qq'} =
\pmatrix{-(2l)^2f_1&\Lambda_3&-\Lambda_2\cr-\Lambda_3&\Omega^2&\Lambda_1\cr\Lambda_2&-\Lambda_1&\Omega^2}$$

 Inverting this matrix and following the procedure to read off the new
fields we obtain the new solution.

 Since ${{\cal{E}}}_{im}=0,\ f^{-1}\ $ represents ${\hat g}
+{\hat B}$ $\big($ For constant $r$ surfaces the $dr$ terms remain
unchanged$\big)$.

so

\ben
\ \hat{g} &=& {1\over\Delta}
\bordermatrix{&\bar\partial \Lambda_1
&\bar\partial \Lambda_2 &\bar\partial \Lambda_3\cr
&\Omega^4 + \Lambda_1^2 & \Lambda_1\Lambda_2&\Lambda_1\Lambda_3\cr
&.&\Lambda_2^2-(2l)^2f_1\Omega^2&\Lambda_2\Lambda_3\cr
&.&.&\Lambda_3^2-(2l)^2f_1\Omega^2}_{sym}\nonumber \\
\hat{B}&=&{1\over\Delta}
\pmatrix{0&-\Omega^2\Lambda_3&\Omega^2\Lambda_2\cr
.&0&(2l)^2f_1\Lambda_1\cr
.&.&0}_{a-sym}\nonumber \\
\Delta&=& \Omega^2\Bigl( \Lambda_2^2 + \Lambda_3^2 \Bigr) -
(2l)^2f_1\Bigl(\Omega^4+\Lambda_1^2\Bigr) \nonumber \\
\Phi&=& \log(\Delta)\, .
\een

 Since SU(2) is semisimple, these massless modes are a solution of the
$\beta$-function
equations \ref{betafns} \cite{steve}. This has been explicitly checked.
Comparison with the Schwarzschild $SO(3)$ dual suggests that we make
the change of
coordinates:-
$$\Lambda_2=y\cos\alpha , \Lambda_3=y\sin\alpha$$
$$ \Lambda_1^2=(2l)^2t^2$$
yielding the solution $TN_{SU(2)}$.

\ben \label{new}ds^2&=&-{dt^2\over f_1}+{dr^2\over f_1} +
{(2l)^2\over\Omega^2}dy^2-{{\big(\Omega^2ydt+(2l)^2 f_1 t
dy\big)^2}\over{\Delta f_1 \Omega^2}}
+{{f_1\Omega^2(2l)^2}\over\Delta}y^2d\alpha^2\nonumber \\
 B&=&{{\Omega^2y^2(2l)}\over\Delta}d\alpha\wedge
dt+{{(2l)^3f_1yt}\over\Delta}d\alpha\wedge dy \nonumber \\
\Phi&=&\log(-(2l)^2\Delta)\nonumber \\
\Delta&=&f_1\big(\Omega^4+(2l)^2t^2\big)-\Omega^2y^2\,
\een

\bigskip
{\bf{Periodification of coordinates on the dual manifold}}

\medskip

The metric defined by \ref{new} is invariant under
$$t\rightarrow-t\ \ \mbox{and} \ y\rightarrow -y.$$

 These points may be identified to give an orbifold. However,
such an identification changes the sign of the $B$-field. Given a coordinate
basis, if we take
reciprocal periodicities then the solutions are identical points in
the string moduli space. In the case where we take the duals over a
non-coordinate basis it is not so clear how to relate the
solutions. In any case, as in Taub-NUT space, the periodicities of the new
coordinates should be defined
to remove any conical singularities in the new solution. The solution
will then satisfy the $\beta$-function equations everywhere.

\bigskip

 Finally, we see that this solution is the same as the solution \ref{HSO3T}, up
to a constant factor in the $B$-field, hence we find that the abelian and
non-abelian duals commute with each other:

\be
TN_{SU(2)}=TN_{t,SO(3),\Lambda_0}=H_{SO(3),\Lambda_0}\,.
\ee
 Of course this also implies that the $\alpha$ dual of
$TN_{SU(2)}$ gives us $H_{SO(3)}$

\subsection{$O(1,1)$ duality of Taub-NUT}

 As mentioned previously, we may extend the $t\rightarrow 1/t$
discrete duality of Taub-NUT to an $O(1,1)$ group of dualities.
The general solution obtained is \cite{jm}
\ba
ds^2&=&-{f_1\over f_2}(dt+(x+1)l\cos\theta
d\phi)^2+{\frac{dr^2}{f_1}}+(r^2+l^2)(d\theta^2+\sin\theta^2d\phi^2)\nonumber
\\
f_2&=&1+(x-1){{mr+l^2}\over{r^2+l^2}}\nonumber \\
\Phi&=&\Phi(r)\nonumber \\
B_{\phi t}&=&(x-1)\cos\theta \times (r-\mbox{terms})\nonumber \\
x^2&\geq1&\,
\ea
of which the H-monopole is the $x=-1$ case and Taub-NUT is the $x=1$ case.

\bigskip

It is interesting to note that either the $dtd\phi$ metric cross terms or the
$B$-field vanish only for the $x=\pm 1$ cases. In these two
cases we are able to perform the duality transformation for the
non-abelian spherical symmetries. In the case $|x|\geq 1$ we are
unable to factor out the $SO(3)$ dependence from the matrix $Q$ to
give a real $\cal{E}$. This creates problems with performing the
duality transformation, and seems to
single out the extreme cases of small and large radius
compactification as being special in some way.

\subsection{Equivalence of Solutions}

We stress that all the solutions we have presented are completely equivalent as
low energy string theory solutions since the gauged subgroups are
either abelian or semisimple. There are interesting connections between
the new solutions, which may be represented diagrammatically as follows:

\ben \mbox{TN}&\stackrel{SU(2)}\longrightarrow&\mbox{TN}_{SU(2)}\nonumber \\
\uparrow&  \ \ &\uparrow\nonumber \\
t,\Lambda_0& \ \ &\Lambda_0,\alpha \nonumber \\
\downarrow& \ \ &\downarrow \nonumber \\
\mbox{H}&\stackrel{SO(3)}\longrightarrow&\mbox{H}_{SO(3)}\,
\een

Thus the duality procedure closes, the abelian duality commutes with the
non-abelian duality transformation at least for the case of Taub-NUT,
and the extreme points of the abelian
duality seem to have special significance with respect to the abelian
duality.

\section{Properties of $TN_{SU(2)}$}\label{proptn}
\news

We now investigate the solution obtained after performing the
non-abelian transformation on Taub-NUT space. Despite the complicated
local behaviour of the solution, it has some global properties which
may be understood from a geometrical point of view (see the figure). The global
structure of the solution clarifies the
relationship between spacetime singularities and fixed points of
isometry groups.
\bigskip

 {\bf Signature}

 The determinant of the metric is
\ben\det(g)&=&-y^2(2l)^4(r^2+l^2)^2\left[\Delta^{-2}\right]\nonumber \\
\Delta&=&f_1\big(\Omega^4+(2l)^2t^2\big)-\Omega^2y^2\,
\een
This is only zero for non-singular regions when $y=0$. The determinant
never becomes positive which, coupled with an examination of the behaviour of
the
coordinate surfaces, shows that we still have a completely Lorentzian
spacetime.

\bigskip

{\bf Singularities}

 The metric has curvature singularities (singular Riemann curvature
terms) when:-

$i$ \ \ \  $\Omega^2=r^2+l^2=0$

$ii$ \ \   $f_1=1-2\Big({{mr+l^2}\over{r^2+l^2}}\Big)=0 \iff r=r_{\pm}$

$iii$ \ $\Delta=0$

 The surface $\Delta=0$ has a strictly {\it spacelike} normal.
\medskip

{\bf Geometry}

 There are four separate regions which need to be considered

\medskip
{\ I: $\Delta>0$,  $r>r_+$ $(f_1>0)$}

\medskip
 For large $t$ and $r$ the spacetime is non-singular and
for small $y$ approaches  the metric of a 2-dimensional H-monople-type throat
region where the $y$
and $\alpha$ dimensions are squeezed:

\ben &&ds^2 \rightarrow -{1\over f_1}dt^2+{1\over f_1}dr^2 + O\left({1\over
{r^2}}\right)(dy^2,dydt,d\alpha^2) \nonumber \\
&&B\rightarrow {2ly^2\over{f_1r^2}}d\alpha\wedge dt +O\left({1\over
{r^4}}\right)dy\wedge dt \,
\een

 This gives us the interpretation
of $t$ as the asymptotic time.

\smallskip

 For a given $(t,r)$ there is a singularity, for large enough
$y=y_{sing}$, where $\Delta$ vanishes.  For an observer at fixed $r$
and $y$, $y_{sing}$ increases as $t^2$:-
\ben
y_{sing}^2&=&f_1\left(\Omega^2+\frac{(2l)^2t^2}{\Omega^2}\right)\nonumber \\
&\sim&\frac{t^2}{r^2}\,
\een

 Thus the singularity accelerates away from such observers, in the
manner of a particle in a constant potential. In this region when
$y^2>f_1\Omega^2$, surfaces of constant $t$ have a spacelike
normal. Thus there are no timelike curves with fixed
$(r,y,\alpha)$ and we have an {\it ergoregion}.
In self-dual euclidean Taub-NUT space, geodesics can be thought of as
describing two monopoles interacting via a Coulomb type potential
\cite{gm}. This seems to be a similar property to the acceleration
described above, and is is possible that the dual space retains some
sort of
monopole interpretation.

The solution also displays some of the properties of the
$C$-matrics, $\cite{kinn}$, in this region. The general form of the solution is
similar to the
C-metrics, and the acceleration property of the horizon is also reminiscent
of such metrics.
\bigskip

{{II:$\Delta<0,\ r_-<r<r_+\  (f_1<0)$}}

\medskip
 This region is completely non-singular. The timelike variable is give
by $r$ and all other variables are spacelike, as in region II
of Taub-NUT.

\bigskip
{{III: $\Delta>0,\ \  r<r_- \ (f_1>0)$ }}

\medskip
This region has similar properties to region I. In the special case
where $m=0$ the two regions I and III are isometric and may be identified.

\bigskip
{{IV: $\Delta<0$ ,\ $r>r_+ \ (f_1>0)$}}
\medskip

In this region $\alpha$ is timelike and the normal to sufaces of
constant $\alpha$ are timelike. Therefore, since $\alpha$ is an angular
variable we have closed timelike curves along orbits of constant $t,r,y$.

\bigskip

 {\bf{Comparison with Taub-NUT}}

 The geometric interpretation of
Taub-NUT space and the metric of the dual are similar: both can be
thought of as two cones
free to rotate on an internal axis separated by a non-singular
region. The nature of the dual geometry may be intuitively understood by
considering the fixed points of
the Taub-NUT symmetry space.

i) The duality procedure
fixes the position of the Taub-NUT cone in $\rr ^3$. This position
corresponds to the choice of gauge.

ii) Secondly, the boundary becomes
singular. This corresponds to the invariance of Taub-NUT under the
right action of shifts in $\psi$. Notice that in region II of Taub-NUT
there are no boundaries (fixed points of the $SU(2)$ action), hence region II
of the dual is entirely
regular.

 These are generic feature of duality procedures:
fixed points of the action of the isometry group on the original manifold
become singular
points in the dual model.
This is demonstrated in the Schwarzschild dual where we have a
line of singularities corresponding to the $\theta=0$ line of fixed points of
the $SO(3)$ isometry, and the abelian T-dual of two dimensional Minkowski space
in
polar coordinates, where
a point naked singularity is introduced, corresponding to the fixed point
$\theta=0$.

\section{Conclusion}
\news

We have found dual descriptions, both abelian and non-abelian, of
Taub-NUT space in string theory. The dualities commute at least for
the case of Taub-NUT space.  The resulting spaces are rather complicated and
lose their non-abelian symmetries. It seems an important task to study
the nature of the dual spaces, which have
properties reminiscent of some point particle spacetimes, such as
those of $C$-metrics and monopole spacetimes. Although the original and dual
spaces seem very different, they are equivalent
as string theory vacua and in order to understand the full string
theory moduli space, the relationship between these string
backgrounds needs to be better understood.

The framework of non-abelian dualities may be applied to ${\it{any}}$
string solution with a background which admits non-commuting Killing
vectors. This has an extremely wide area of applicability: many string
solutions, when considered as full ten dimensional solutions, have a pure
spherical symmetry arising from extra flat dimensions, such as in
plane waves. These extra
dimensions may be dualised over to give
complicated solutions of the form \ref{xxxx}. Perhaps in string
theory, spherical symmetry is not
as simple as it seems from a particle point of view.

At the very least non-abelian duality points to an equivalence of
solutions in string theory beyond those given by abelian target space
duality and the
diffeomorphisms of
general relativity.

\bigskip
I would like to thank Malcolm Perry for suggestions and help.
\newpage

\newpage
\vfill\eject

\end{document}